\documentclass[acmsmall,manuscript,screen]{acmart}
\usepackage{rotating}
\usepackage{enumitem}
\usepackage{hyperref}
\usepackage{array}
\usepackage{booktabs}
\AtBeginDocument{%
  }

\setcopyright{acmlicensed}
\copyrightyear{2024}
\acmYear{2024}
\acmDOI{}

\acmISBN{}

\begin{document}
\newcolumntype{L}[1]{>{\raggedright\arraybackslash}p{#1}}

\title{WARNING This Contains Misinformation: The Effect of Cognitive Factors, Beliefs, and Personality on Misinformation Warning Tag Attitudes}

\author{Robert Kaufman}
\email{rokaufma@ucsd.edu}
\affiliation{%
  \institution{University of California, San Diego}
  \streetaddress{9500 Gilman Dr}
  \city{La Jolla}
  \state{California}
  \country{USA}
  \postcode{92093}
}

\author{Aaron Broukhim}
\email{aabroukh@ucsd.edu}
\affiliation{%
  \institution{University of California, San Diego}
  \city{La Jolla}
  \state{California}
  \country{USA}
}

\author{Michael Haupt}
\email{mhaupt@ucsd.edu}
\affiliation{%
  \institution{University of California, San Diego}
  \city{La Jolla}
  \state{California}
  \country{USA}
}

\renewcommand{\shortauthors}{Kaufman et al.}

\begin{abstract}
Social media platforms enhance the propagation of online misinformation by providing large user bases with a quick means to share content. One way to disrupt the rapid dissemination of misinformation at scale is through warning tags, which label content as potentially false or misleading. Past warning tag mitigation studies yield mixed results for diverse audiences, however. We hypothesize that personalizing warning tags to the individual characteristics of their diverse users may enhance mitigation effectiveness. To reach the goal of personalization, we need to understand how people differ and how those differences predict a person's attitudes and self-described behaviors toward tags and tagged content. In this study, we leverage Amazon Mechanical Turk (n = 132) and undergraduate students (n = 112) to provide this foundational understanding. Specifically, we find attitudes towards warning tags and self-described behaviors are positively influenced by factors such as Personality Openness and Agreeableness, Need for Cognitive Closure (NFCC), Cognitive Reflection Test (CRT) score, and Trust in Medical Scientists. Conversely, Trust in Religious Leaders, Conscientiousness, and political conservatism were negatively correlated with these attitudes and behaviors. We synthesize our results into design insights and a future research agenda for more effective and personalized misinformation warning tags and misinformation mitigation strategies more generally.
\end{abstract}




\maketitle

\section{Introduction}
Misinformation, or "fake news", is a long-standing and pervasive issue with many adverse implications. Affecting a wide range of domains, misinformation has been tied to lower vaccine usage during the COVID-19 pandemic \cite{bridgman2020causes}, climate change denial \cite{treen2020online}, and election interference \cite{guess2018selective} among other harmful effects. Social media platforms uniquely contribute to the spread of misinformation due to their large user bases, under-moderation, and the network effects of virality \cite{Gillespie2020ContentMA, Singhal2022SoKCM, Barber2018ExplainingTS}. Since the rise of social media - where 62\% of people get their news \cite{gottfried2016news} - the spread of misinformation has dramatically increased \cite{vosoughi2018spread}. The World Health Organization (WHO) has gone so far as to declare an "infodemic" characterized by the intentional spreading of inaccurate information to undermine public health efforts \cite{world2022managing}. 

As the problem of misinformation continues to grow, it is critical to develop effective mitigation methods that can work at scale. Several approaches have been taken to prevent false belief formation from misinformation as well as curb misinformation spread, each approach depending on the ubiquitousness of the case. Early algorithmic intervention – often using artificial intelligence or crowdsourcing – can be highly effective at moderating content visibility and preventing misinformation from spreading in the first place \cite{Kaufman2022WhosIT, Gillespie2020ContentMA, kim2019homogeneity, roitero2020can, roitero2020covid}. There are far fewer options for misinformation that has already become pervasive on social platforms – i.e. “gone viral”. For these cases, tagging potentially misleading content with misinformation warning labels may be effective \cite{clayton2020real}, but warnings must be convincing and trustworthy to the viewer in order to be taken seriously. This is a tall order, as different people perceive and act on information in different ways \cite{Kaufman2022WhosIT, pennycook2018prior, lees2022twitter, roets2017fake}. If done correctly, warning tags present a potentially transparent \cite{Seo2019TrustIO}, prudent \cite{Saltz2020EncountersWV}, and effective \cite{Martel2023MisinformationWL} means of intervention that prevents the spread of misinformation.

\begin{figure}[h]
  \centerline{\includegraphics[width=0.5\textwidth]{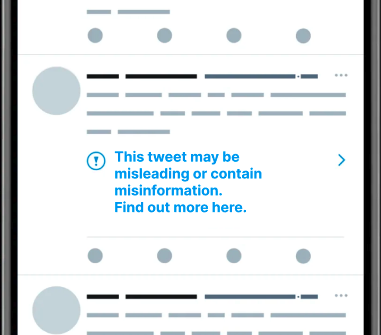}}
  \caption{Warning tag example shown to participants (from \cite{warningTagExIMG})}
  \label{fig:warning-tag}
  \Description{A visual representation of a warning tag.}
\end{figure}

In this paper, we focus on warning tags for misinformation on social media as these are a front-line defense to prevent the proliferation of fake news at the point of human interaction and information uptake. We define "warning tags" as communicative actions that are spatially and temporally attached to a piece of information, news story, or source. The purpose of the tag is to alert the viewer that the content they are about to view is potentially unreliable, false, and/or encouraging them to exercise caution before accepting or sharing the information \cite{clayton2020real}. This is a common method for misinformation prevention on social media, and has been used on popular social media platforms such as X (formerly Twitter) and Facebook \cite{Pennycook2019TheIT, FBabout}. Figure \ref{fig:warning-tag} shows an example from X shown to participants in the present study. Despite the wide adoption of tags across social media platforms, past studies show mixed results with regards to the effectiveness of tags in combating misinformation belief and spread. Differences in efficacy are often attributed to political orientation and cognitive ability \cite{lees2022twitter, pennycook2018prior, roets2017fake}. While these factors may predict tag effectiveness in some cases, we posit that additional characteristics may provide further insight into what types of tags work best and for whom. 

An intuitive method to mitigate group differences in tag efficacy is to tailor communications to the diverse needs of the audience. In practice, few studies have explored this possibility \cite{Saltz2020EncountersWV}, and no major social media platform – to our knowledge – has attempted to implement personalized tags in practice. A lack of personalization limits the effectiveness of the labels due to individual differences in perception and information-related behaviors \cite{Kaufman2022WhosIT, pennycook2018prior, lees2022twitter, roets2017fake}. Thus, one-size-fits-all approaches to prevent misinformation uptake and spread will fail if they don’t harmonize with the specific needs and motivations of a particular individual. In the case of misinformation, tags warning of potential falsehoods need to be tailored to address the perceptions of the specific recipient, while online interventions designed to convince someone away from spreading falsehoods need to be met at the motivating source. We posit that the current one-size-fits-all implementation of misinformation tags can benefit from modifications that account for personal and contextual flexibility. Personalization and information tailoring have been topics of discussion in many other domains where people need to make sense of AI-based decisions \cite{holzinger2019causability, wang2019designing, schneider2019personalized}, ranging from knowledge systems \cite{lim2009and}, healthcare \cite{kaufman2022cognitive, kaufman2023explainable}, image classification \cite{pazzani2022expert, soltani2022user}, and autonomous driving \cite{kaufman2024effects, kaufman2024developing}. The need for tailored messaging is also discussed in user persona work that accounts for multiple factors such as psychological traits, situational circumstances, and demographics to produce nuanced depictions of public response towards health issues \cite{haupt2022psychological, massey2021development, huh2016personas, alsaadi2021use}. In a similar vein, effective interventions for misinformation, especially ones that are visible to users, need to account for the diversity of user reactions. However, there is little known about factors that influence user attitudes towards tags, limiting efforts to develop customized interventions. We aim to fill this gap with the present study.

The work presented in this paper follows up on previously published work by Kaufman et al. on personal factors impacting misinformation assessment, where the results showed that individual cognitive and information assessment traits and personal beliefs can be used to predict tweet misinformation assessment accuracy \cite{Kaufman2022WhosIT}. In this study, we use datasets from two diverse sample populations (undergraduates and Amazon Mechanical Turk workers) to predict attitudes towards misinformation warning tags from similar traits. We measure cognitive and personality traits, political orientation, and institutional trust with the purpose of highlighting how individual differences impact attitudes toward misinformation warning tags. The purpose of this work is to establish a foundation for personalized warning tags and provide a deeper understanding of factors that need to be accounted for to effectively mitigate misinformation uptake and spread at the individual level.

Our results indicate that Big Five Inventory (BFI) personality trait Openness, political orientation, and Trust in Medical Scientists were the most influential predictors for favorable attitudes towards tags among our undergraduate sample. Among a more general population sample collected via Amazon Mechanical Turk (MTurk), we found that the dispositional trait Need For Cognitive Closure (NFCC) showed the strongest association with positive warning tag attitudes. BFI Conscientiousness, political orientation, Trust in Medical Scientists, and trust in religious figures were also influential factors on tag attitudes. 
\vspace{1em}

In sum, the study presented here contributes:
\begin{enumerate}
  \item An analysis of the individual differences contributing to warning tag attitudes and self-described behaviors.
  \item A comparison of two widely-used human-computer interaction research populations: Amazon Mechanical Turk (MTurk) and undergraduate study participants (SONA) in the context of misinformation warning tags.
  \item Insights and design implications for personalized and user-centric warnings tags that can more effectively mitigate the spread of misinformation among diverse populations.
\end{enumerate}

Taken in full, this paper provides insight into how personal factors impact a person’s attitude towards misinformation warning labels, an important and necessary first step towards developing tailored interventions that are effective not just for some, but for all.

\section{Related Work}
\subsection{Individual Differences in Misinformation Vulnerability and Propagation}
Several prior studies have examined misinformation vulnerability and spreading behavior. In this section, we focus on \textit{who} is most vulnerable and likely to spread misinformation and use this as a basis to inform our study on individual differences in attitudes towards tag interventions.

Increasing reliance on social media for news contributes to the rapid dissemination of misinformation on social media \cite{gottfried2016news, vosoughi2018spread}. Since 2016, foreign agents have leveraged social media to attempt to influence the United States presidential election in favor of their preferred candidate \cite{8508646}, damaging voter agency and potentially swaying election outcomes against the interest of the electorate. Following this election, in which the term "fake news" became pervasive in popular culture, the COVID-19 pandemic showed the influence of misinformation via debate around vaccine efficacy. A study by Loomba et. al showed a negative correlation between exposure to misinformation and intent to vaccinate against the virus, with potentially detrimental health outcomes \cite{loomba2021measuring}. Importantly, this study was also one of the first to highlight individual differences in misinformation susceptibility and behavioral intention to vaccinate. They found that people of different races, religious backgrounds, and genders demonstrated different levels of susceptibility to misinformation. 

In the United States, prior work has consistently highlighted political differences in misinformation vulnerability \cite{Levin2022DeterminantsOC} and sharing behavior \cite{haupt2021characterizing, haupt2021identifying}. Past studies show that people who identify as politically conservative share more misinformation and may be less accurate at identifying misinformation than than those who identify as liberal \cite{Guess2019LessTY, pretus2023role, Kaufman2022WhosIT}. The divide is not absolute, however: both liberals and conservatives are more likely to share misinformation that positively reflects their own ideological groups \cite{Pereira2018IdentityCD} and seek out information that confirms their previously held beliefs \cite{Druckman2019TheEF}. Partisanship also plays a role in attitudes toward misinformation interventions: Saltz et al. \cite{Saltz2021MisinformationIA} found heavily-leaning liberals supported a variety of online misinformation interventions, while conservatives opposed interventions outright. We seek to build on this past work by exploring how political orientation may impact a person’s attitudes towards tag mitigation solutions in conjunction with other personal traits.

There is a mounting archive of evidence suggesting that the informational sources a person trusts may impact their susceptibility to misinformation, including interpersonal connections, institutions, and news outlets. Intuitively, trusting experts – particularly medical experts – has been shown to decrease misinformation susceptibility, while trust in family, friends, and less reliable sources increases susceptibility \cite{Kaufman2022WhosIT, callaghan2020correlates, motta2018knowing}. Surprisingly, individuals with high trust in science may be more susceptible to misinformation that contains pseudo-scientific content \cite{OBrien2021MisplacedTW}. Divergent from trust in science, Trust in Religious Leaders has been associated with belief in misinformation \cite{pennycook2021psychology}. It's worth noting that Jasinskaja-Lahti et el. highlights a distinction between individuals who self-categorize as religious and individuals who endorse religious worldviews, stating that endorsement of religious worldviews may be a better predictor of susceptibility to conspiracy theories \cite{JasinskajaLahti2019UnpackingTR}. Prior work by Kaufman et al. found crowdworkers with less Trust in Religious Leaders to be more effective at labeling misinformation \cite{Kaufman2022WhosIT}. Trust characteristics may underlie political differences in misinformation vulnerability, sharing behavior, and attitudes \cite{Agley2020MisinformationAC}. We seek to understand how trust in informational sources including news media, elected officials, and medical scientists and religious leaders may impact the viability of warning tags for misinformation mitigation.

Prior studies have shown that cognitive, information assessment, and personality traits may impact misinformation vulnerability and spreading behavior \cite{Kaufman2022WhosIT}. Kaufman et al. found that respondents with high Cognitive Reflection Test (CRT) scores, Conscientiousness, and Trust in Medical Scientists were more aligned with experts at an information assessment task while respondents with high Need for Cognitive Closure (NFCC) and those who lean politically conservative were less aligned with experts.

Often measured by the Cognitive Reflection Test (CRT), tendency for reflective thinking has been found to be correlated with misinformation sharing and intervention attitudes. CRT performance is associated with the ability to override incorrect analytical intuitions through reflection. Pehlivanoglu et al. found positive correlations between CRT score and an individual's ability to identify misinformation \cite{Pehlivanoglu2020TheRO}. Another study showed individuals with lower CRT scores were more willing to overclaim knowledge and had a general tendency to be more receptive to "pseudo-profound bullshit" \cite{pennycook2020falls}. CRT has also been shown to be positively associated with trust in medical scientists \cite{Larsen2023TheIO}. In the present study, we include CRT as a predictor for tag attitudes.

Other information assessment traits, such as a person’s Need for Cognitive Closure (NFCC), have been shown to impact a person’s vulnerability to misinformation \cite{bessi2015science, pica2014role}. Need for Cognitive Closure (NFCC) is a measure of one's desire for order, predictability, and discomfort with ambiguity \cite{webster1994individual}. Prior work has shown NFCC to be negatively correlated with misinformation identification accuracy \cite{Kaufman2022WhosIT}. Other studies have shown that individuals with high NFCC may be more likely to spread misinformation on online social media networks, with the effect attributed to avoidance behavior when asked to provide evidence for beliefs \cite{bessi2015science}. The effect of NFCC in other information evaluation domains has been contested \cite{de2020investigating}. We seek to provide clarity on the impact of NFCC on warning tag attitudes, as warning tags may impact the level of ambiguity for online information assessment.

Prior work has established connections between personality and misinformation vulnerability \cite{Kaufman2022WhosIT, Lai2020WhoFF} and spreading behavior \cite{Indu2021ASR, Alvi2020INFORMATIONP, lawson2022pandemics}. The Big Five Inventory (BFI) Conscientiousness scale, for example, measures an individual's tendency to be orderly, cautious, and self-disciplined. Conscientiousness has been associated with better news discernment \cite{calvillo2020political}, and individuals low in Conscientiousness may be more likely to share fake news \cite{lawson2022pandemics}; although, later research refutes these claims \cite{Lin2023ConscientiousnessDN}. BFI Openness to Experience measures a person's curiosity, receptivity to new experiences, and imagination. Some reports find that individuals high in Openness may be more vulnerable to misinformation \cite{Ahmed2022PersonalityAP}. However, there is also conflicting evidence observed with this BFI trait: prior work demonstrates a negative association between Openness and belief in myths \cite{swami2016believes} and a positive association with news discernment \cite{calvillo2020political}. BFI agreeableness has been associated with users who validate news prior to sharing \cite{sampat2022fake}, and has been included in our study as well. In the present work, we assess whether personality influences tag attitudes and self-described behaviors, as these dimensions may impact how information is evaluated and perceptions of interventions.

Despite all of these studies assessing the impact of individual factors on misinformation vulnerability and spread, very little work has focused on the impact of personal factors on misinformation mitigation solutions like warning tags \cite{Grady2021NeverthelessPP} specifically. No prior work, to our knowledge, assess the impact of individual differences at the level of cognitive, informational, or personality traits or misinformation identification accuracy on tag attitudes. In this work, we seek to fill these gaps.

\subsection{Misinformation Warning Tags as Interventions}
Due to the ubiquitous nature of misinformation spread online, deployment of mitigation solutions that can work at-scale is vital. Prior studies have found the efficacy of tags and other intervention methods to prevent belief in misinformation is complex and multi-faceted.

Attempts to curb the sharing of misinformation often involve moderation. Determining what content to moderate is a difficult problem within itself: algorithms that can work swiftly at scale typically come at the cost of nuance \cite{Gillespie2020ContentMA}. Recruiting a human to handle moderation edge-cases can help \cite{Lai2022HumanAICV}, but this approach assumes underlying algorithms can identify edge cases accurately. Further, human-identified misinformation introduces the partiality of a human into the moderation process \cite{Haimson2021DisproportionateRA}. Some users may be skeptical of warning tag accuracy depending on if the correction comes from a community member or an algorithm \cite{Jia2022UnderstandingEO} and manly believe tags from humans are more biased than tags from algorithms \cite{Wang2020ModeratingUU}. As a result, AI-based tagging of potential misinformation and AI-generated labeling of online posts is seen as a potential soft-moderation middle ground that can help curb misinformation spread and vulnerability without completely removing the agency of the reader. Deployment efforts have been made by a number of social media platforms, including Facebook and X (formerly Twitter) \cite{X}.

While the effectiveness of warning labels for misinformation has seen generally positive results, there remains some debate on their efficacy. As a whole, misinformation tags have been generally shown to have a modest, positive effect on reducing belief in misinformation \cite{Pennycook2019TheIT, clayton2020real, kim2019homogeneity, Mena2020CleaningUS}. Initial reports of a "backfire" or "boomerang" effect that claim misinformation corrections might actually increase misperceptions \cite{Nyhan2010WhenCF} have since been refuted \cite{Wood2019TheEB}, however, additional works have found that domain may play a role in correction effectiveness \cite{Chan2023AMO}. For example, a meta-analysis by Chan et al. \cite{Chan2023AMO} found correction attempts similar to misinformation tag labels are more successful in scientific domains outside of health.

We hypothesize one of the reasons for prior mixed results on the efficacy of tags is due to the complex nature of misinformation vulnerability, spreading behavior, and attitudes towards mitigation efforts. Misinformation vulnerability and spreading behavior is affected by a person’s psychological faculty \cite{keersmaecker2017FakeNI}, frequency of experience with the information \cite{pennycook2018prior}, personal beliefs \cite{Jia2022UnderstandingEO}, and socioeconomic status \cite{Wang2022AlternativeSU}. Intervention-centered factors include the phrasing of the intervention \cite{clayton2020real, Ecker2010ExplicitWR} and prevalence of alternative information during intervention \cite{Kan2021ExploringFT}. Though not a complete list, these findings demonstrate the complexity of evaluating the efficacy of misinformation tags. In the present study, we hope to illuminate individual factors which may impact tag efficacy.

Given the wide range of individual factors which may impact a person's attitudes towards tags, it is unsurprising that prior research on warning label attitudes has resulted in mixed findings. As noted above, Saltz et al. found conservatives oppose labeling interventions while liberals supported labeling \cite{Saltz2021MisinformationIA}. Additionally, both conservatives and democrats attributed perceived labeling errors to bias in human judgments against their beliefs rather than accidental algorithmic or human mistakes. Other works support the notion that who provides the warning label matters. Jia et al. found warning labels provided by an algorithm, community, or third-party fact-checker were trusted by Democrats regardless of post ideology while only algorithmic labels impacted Republican's belief of false conservative news \cite{Jia2022UnderstandingEO}. All intervention methods were effective for Republicans when the content was liberal-leaning. 

Though it is clear that there are strong individual differences in the efficacy of tags, little work has been conducted on the personalization of warning tags. In a related study, Saltz et al. identifies significant variance in individual preferences for warning tags \cite{Saltz2020EncountersWV}. When asked about their attitudes towards warning tags, a plethora of concerns arose for users. Some users felt patronized by the social media platforms, thought the platforms were being too political, found the labels insufficient, or preferred that the posts be removed altogether. A recent attempt at personalizing moderation in the context of misinformation, Pretus et al. presents posts in a more culturally aligned way to the user's identifying political party, finding that users are more receptive to posts that align with these political norms, especially in polarizing contexts, providing "initial evidence that identity-based interventions may be more effective than identity-neutral alternatives for addressing partisan misinformation" \cite{Pretus2024TheMC}. The implication is that personalization may be an effective direction to increase the efficacy of mitigation efforts like tags. In order to get there, we need to know what individual differences matter.

Though prior work largely attributes discrepancies in warning tag efficacy to political affiliation, there is increasing evidence that other characteristics better account for an individual's misinformation susceptibility and sharing behaviors \cite{Kaufman2022WhosIT, piksa2022cognitive, sun2024shares, lawson2022pandemics}. While this work finds that Politics is correlated with attitudes towards warning tags, factors such as CRT, NFCC, BFI Conscientiousness, Trust in Medical Scientists, and Trust in Religious Leaders were also important factors that account for attitudes towards warning tags. Such results are important to design and deploy misinformation warning labels that can meet the needs of diverse groups.

\section{Method}
Participants from Amazon Mechanical Turk (MTurk) and SONA - an undergraduate population - filled out survey scales to determine their Political affiliation, Trust in Medical Scientists, BFI scores, CRT, and NFCC. The resulting scores were then analyzed with respect to the attitude scale found in Tables \ref{tab:bivCorSONA2022} and \ref{tab:bivCorMTURK2021}. Correlations were evaluated to determine what individual characteristics best predict respondent attitudes towards warning tag labels. 

\subsection{Data Collection}
The first set of respondents were recruited from SONA (n = 112), a pool of undergraduate students from a large university in the United States. The mean age of the SONA respondents was 21.6, of which 71.4\% identified as female. For the MTurk sample (n = 132), the mean age was 37.3 years old and 42.4\% identified as female. All respondents were anonymous. Respondents who failed an attention check and/or finished under the 10th percentile of completion time (<4.5 minutes, median completion time = 24 minutes) were excluded from analysis. Compensation varied between sources: MTurk workers were compensated financially at standard survey-taking rates, while SONA respondents were given course credits. Additional variables included in the SONA sample to further examine effects from personality factors (i.e., openness, agreeableness, neuroticism, and extroversion). See Appendix Table \ref{tab:meanCompPlat} for detailed descriptions of the samples used in the present analysis. 

\subsection{Variable Descriptions}
To measure attitudes and self-described behaviors towards misinformation warning tags - the main outcomes of this study - a tag attitude scale was adapted from existing scales used to assess attitudes towards explainable AI explanations (these provide justifications for AI-decisions, similar to algorithmically-generated tags) \cite{shin2021effects}. The tag scale consisted of 9 individual 5-point likert-scale items ranging from strongly disagree to strongly agree. Participants first saw the statement "When I see a social media post tagged as potential misinformation..." followed by each item. An image of a warning tag within a newsfeed was also shown as an example to respondents (See Figure \ref{fig:warning-tag}). An aggregate (overall) score was calculated by averaging across items. The full scale can be viewed in Table \ref{tab:bivCorMTURK2021}. A higher scale score indicates more favorable attitudes towards misinformation tags. The first item “I usually ignore the tag” was reversed coded since it expresses negative sentiment towards tags. Tag attitude scale items showed acceptable consistency (Cronbach’s alpha = 0.752) and was normally distributed (Shapiro-Wilk = 0.176) within the SONA sample. For the MTurk sample, the tag scale showed less consistency (Cronbach’s alpha = 0.546) and was not normally distributed (Shapiro-Wilk p < 0.01). 

To assess individual differences, the following scales were used:
\begin{itemize}[nosep, topsep=0pt, partopsep=0pt]
    \item \textit{Political orientation} - One question asking participants to identify their political beliefs, with 1 = Strong Democrat/Liberal to 6 = Strong Republican/Conservative.
    \item \textit{Trust in Medical Scientists} - One 4-item scale adapted from the 2019 Pew Research Center’s American Trends Panel survey \cite{funk2019trust} asking respondents "How much confidence, if any, do you have in each of the following to act in the best interests of the public?". The institutions asked about were elected officials, news media, Medical Scientists, and religious leaders with response options ranging from 1 = No confidence at all to 4 = A great deal. 
    \item \textit{Big Five Inventory} - One 41-item scale from the Big Five Inventory (BFI) was used to evaluate participants across the personality dimension of Extroversion, Neuroticism, Agreeableness, Openness, and Conscientiousness \cite{john1991big, john2008paradigm, rammstedt2007measuring}. Only the subscale for Conscientiousness was given to participants in the MTurk samples due to length constraints. 
    \item \textit{Cognitive Reflection Test} - The Cognitive Reflection Test (CRT) was employed to assess cognitive reflection using three questions. Each question initially suggests an "intuitive" answer, but arriving at the correct solution requires reflective thought. For instance, Question 1 states: "A bat and a ball together cost \$1.10. The bat costs \$1.00 more than the ball. How much does the ball cost?" The intuitive answer is 10 cents, but the correct answer is 5 cents. A higher number of correct answers indicates a higher CRT \cite{frederick2005cognitive}.
    \item \textit{Need for Cognitive Closure} - A shortened 15-item scale \cite{roets2011item}, adapted from the original 42-item scale \cite{webster1994individual}, was utilized to assess NFCC.
\end{itemize}

Bivariate correlations were run between all trait variables and each individual tag item. Tag items were not normally distributed, therefore the Kendall’s Tau correlation coefficient was used. Correlations were also run between tested trait variables and the composite tag scale. Since the composite tag scale was normally distributed, Pearson’s correlation was used with traits that were also normally distributed while Kendall’s Tau was used for remaining variables. Multiple regression was also run between tested traits and the composite tag scale as the dependent variable. In order to account for potential multicollinearity of independent variables, a Shapley value regression was run to show the variance contributed by each tested trait.

\section{Results} 
First, we present bivariate correlations of the aggregate attitude values and their respective components to individual characteristics (BFI, NFCC, etc.) for each sample. Then, we present multiple regression results for each sample, which include Shapley value coefficients to break down the top factors predicting warning tag attitudes. Finally, the biavariate correlation and mutlivariate regression results are compared.

\subsection{Predicting Tag Attitudes}
Bivariate correlations were run on the SONA sample between tag attitudes items and individual traits as shown in Table \ref{tab:bivCorSONA2022}. Positive correlations were shown between the composite tag attitude scale and BFI traits for Openness (r = 0.40, p < 0.01), Conscientiousness (r = 0.19, p < 0.05), and Agreeableness (r = 0.29, p < 0.01). For demographic variables, the tag attitude scale was positively correlated with stronger identification with being a Democrat (0.36, p < 0.01) and higher Trust in Medical Scientists (r = 0.27, p < 0.01). However, Trust in Religious Leaders was negatively correlated with tag attitudes (r = -0.20, p <.01). All variables that showed significant correlations with the composite attitude scale also showed correlations with multiple individual tag items. While CRT shows no significant correlation with the tag attitude scale in the SONA sample, it was positively associated with the tag item “I am more likely to research the post’s topic in order to evaluate the content myself” (r = 0.18, p < 0.05). 

\begin{table*}
\small
  \caption{Bivariate correlations between tested variables and tag attitude items (SONA)}
  \label{tab:bivCorSONA2022}
  \begin{tabular}{p{2.2cm}|p{1.0cm}|p{1.0cm}|p{1.0cm}|p{1.0cm}|p{1.0cm}|p{1.0cm}|p{1.0cm}|p{1.0cm}|p{1.0cm}|p{1.3cm}}
    \toprule
     & \multicolumn{1}{|L{1.0cm}|}{\textbf{Overall}} & \multicolumn{1}{|L{1.0cm}|}{"I usually ignore the tag"} & \multicolumn{1}{|L{1.0cm}|}{"I am less likely to read the post”} & \multicolumn{1}{|L{1.0cm}|}{“I am less likely to engage with the post through sharing, replying, or liking”} & \multicolumn{1}{|L{1.0cm}|}{“I evaluate the post content based on who the author is”} & \multicolumn{1}{|L{1.0cm}|}{“I am more likely to research the post's topic in order to evaluate the content myself"} & \multicolumn{1}{|L{1.0cm}|}{“I approach the post content with\newline suspicion or \newline hesitancy”} & \multicolumn{1}{|L{1.0cm}|}{“I view the author of the post in a negative way"} & \multicolumn{1}{|L{1.0cm}|}{“I trust that the post is, in fact, misinformation"} & \multicolumn{1}{|L{1.2cm}}{“I appreciate the social media platform's attempt to warn me about potential misinformation"} \\
    \midrule
    CRT & 0.070 & 0.020 & -0.036 & -0.074 & 0.130 & \textbf{0.180*} & 0.111 & 0.010 & 0.000 & 0.065 \\
    NFCC & 0.050 & 0.001 & 0.096 & 0.0270 & -0.025 & -0.030 & 0.063 & 0.050 & -0.060 & 0.0280 \\
    \textbf{Openness} & \textbf{0.400**} & \textbf{-0.243**} & 0.091 & \textbf{0.244**} & \textbf{0.207**} & \textbf{0.280**} & \textbf{0.310**} & 0.078 & 0.130 & \textbf{0.359**} \\
    \textbf{Conscientious} & \textbf{0.190*} & \textbf{-0.199**} & 0.049 & 0.100 & 0.086 & -0.007 & \textbf{0.245**} & -0.020 & 0.060 & \textbf{0.228**} \\
    \textbf{Agreeableness} & \textbf{0.290**} & -0.125 & 0.139 & \textbf{0.173*} & \textbf{0.194**} & 0.061 & \textbf{0.268**} & 0.066 & 0.080 & \textbf{0.256**} \\
    Extroversion & 0.110 & -0.111 & -0.003 & 0.130 & 0.110 & 0.021 & 0.006 & -0.081 & 0.097 & 0.101 \\
    Neuroticism & 0.070 & -0.013 & 0.088 & -0.105 & 0.032 & 0.039 & -0.019 & 0.067 & 0.009 & -0.040 \\
    \textbf{Politics} & \textbf{-0.360**} & \textbf{0.285**} & \textbf{-0.194*} & \textbf{-0.230**} & -0.005 & -0.125 & \textbf{-0.254**} & \textbf{-0.300**} & \textbf{-0.332**} & \textbf{-0.415**} \\
    Trust Elected & -0.020 & 0.139 & 0.110 & -0.024 & 0.068 & -0.001 & -0.140 & 0.160 & -0.007 & -0.111 \\
    Trust News Media & -0.030 & 0.053 & 0.130 & 0.019 & -0.088 & 0.072 & -0.030 & -0.043 & -0.025 & -0.004 \\
    \textbf{Trust Med Sci} & \textbf{0.270**} & -0.102 & 0.125 & \textbf{0.265**} & 0.129 & \textbf{0.252**} & \textbf{0.339**} & 0.043 & 0.040 & \textbf{0.251**} \\
    \textbf{Trust Religious} & \textbf{-0.200**} & 0.127 & -0.144 & -0.114 & -0.015 & -0.094 & \textbf{-0.188*} & \textbf{-0.190*} & -0.061 & -0.162 \\
    Age & -0.050 & 0.067 & -0.051 & -0.028 & -0.092 & -0.043 & -0.114 & 0.104 & -0.006 & -0.082 \\
    \bottomrule
    \multicolumn{11}{l}{Attitude responses prefaced with "When I see a social media post tagged as potential misinformation..."}\\
    \multicolumn{11}{l}{Significance codes: *p < 0.05, **p < 0.01} \\
  \end{tabular}
\end{table*}

Table \ref{tab:bivCorMTURK2021} shows bivariate correlations for the MTurk sample. NFCC was positively correlated with the tag attitude scale (r = 0.447, p < 0.01). Further, Conscientiousness within the MTurk sample was negatively correlated with the tag attitude scale (r = -0.161, p < 0.05). All remaining variables except for Trust in Elected Officials showed no statistically significant effects with the composite tag attitude scale. While CRT was not statistically significant with respect to aggregate tag attitudes, those higher in CRT were less likely to agree with the individual statement "I usually ignore the tag" (r = -0.230, p < 0.01 ). Political orientation did not show a significant correlation with the aggregate attitude scale, however, being liberal was correlated with agreeing with the item "I am less likely to read the post” (r = -0.165, p < 0.05) when seeing a tag for potential misinformation. For trust in new media, only “I appreciate the social media platform’s attempt to warn me about potential misinformation" showed a significant effect (r = 0.182, p < 0.05). Trust in Medical Scientists had three significant correlations with "I usually ignore the tag", "I evaluate the post content based on who the author is", and "I appreciate the social media platform’s attempt to warn me about potential misinformation" despite the aggregate not being significant. Trust in Religious Leaders had significant correlations with "I usually ignore the tag" and "I am more likely to research the post's topic in order to evaluate the content myself". Lastly, the response "I usually ignore the tag" was negatively correlated with age (r = -0.154, p < 0.05).

\begin{table*}
\small
  \caption{Bivariate correlations between tested variables and tag attitude items (MTurk)}
  \label{tab:bivCorMTURK2021}
  \begin{tabular}{p{2.2cm}|p{1.0cm}|p{1.0cm}|p{1.0cm}|p{1.0cm}|p{1.0cm}|p{1.0cm}|p{1.0cm}|p{1.0cm}|p{1.0cm}|p{1.3cm}}
    \toprule
     & \multicolumn{1}{|L{1.0cm}|}{\textbf{Overall}} & \multicolumn{1}{|L{1.0cm}|}{"I usually ignore the tag"} & \multicolumn{1}{|L{1.0cm}|}{"I am less likely to read the post”} & \multicolumn{1}{|L{1.0cm}|}{“I am less likely to engage with the post through sharing, replying, or liking”} & \multicolumn{1}{|L{1.0cm}|}{“I evaluate the post content based on who the author is”} & \multicolumn{1}{|L{1.0cm}|}{“I am more likely to research the post's topic in order to evaluate the content myself"} & \multicolumn{1}{|L{1.0cm}|}{“I approach the post content with\newline suspicion or \newline hesitancy”} & \multicolumn{1}{|L{1.0cm}|}{“I view the author of the post in a negative way"} & \multicolumn{1}{|L{1.0cm}|}{“I trust that the post is, in fact, misinformation"} & \multicolumn{1}{|L{1.2cm}}{“I appreciate the social media platform's attempt to warn me about potential misinformation"} \\
    \midrule
    CRT & -0.005 & \textbf{-0.230**} & 0.031 & 0.013 & 0.001 & -0.115 & 0.091 & -0.038 & -0.134 & -0.054 \\
    \textbf{NFCC} & \textbf{0.447**} & \textbf{0.399**} & \textbf{0.218**} & \textbf{0.320**} & \textbf{0.272**} & \textbf{0.145*} & \textbf{0.297**} & \textbf{0.412**} & \textbf{0.259**} & \textbf{0.285**} \\
    \textbf{Conscientious} & \textbf{-0.161*} & \textbf{-0.214**} & -0.125 & -0.127 & -0.061 & -0.041 & -0.033 & \textbf{-0.167*} & \textbf{-0.180**} & -0.083 \\
    Politics & -0.079 & 0.122 & \textbf{-0.165*} & -0.028 & 0.048 & 0.005 & -0.101 & -0.019 & -0.095 & 0.003 \\
    Trust Elected & 0.084 & 0.133 & -0.005 & -0.053 & 0.117 & 0.123 & 0.085 & 0.053 & 0.096 & 0.124 \\
    Trust News Media & 0.075 & -0.017 & 0.020 & 0.009 & 0.138 & 0.035 & 0.029 & 0.015 & 0.010 & \textbf{0.182*} \\
    Trust Med Sci & 0.109 & \textbf{-0.160*} & -0.061 & 0.146 & \textbf{0.202**} & 0.128 & 0.137 & -0.080 & 0.053 & \textbf{0.206**} \\
    Trust Religious & 0.015 & \textbf{0.212**} & -0.066 & -0.057 & 0.088 & \textbf{0.162*} & 0.069 & -0.054 & 0.013 & 0.100 \\
    Age & -0.077 & \textbf{-0.154*} & -0.114 & -0.108 & 0.032 & -0.019 & -0.087 & -0.016 & -0.129 & -0.036 \\
    \bottomrule
    \multicolumn{11}{l}{Attitude responses prefaced with "When I see a social media post tagged as potential misinformation..."}\\
    \multicolumn{11}{l}{Significance codes: *p < 0.05, **p < 0.01}\\
  \end{tabular}
\end{table*}

Multiple regression model results with Shapley coefficients for the SONA sample are shown in Table \ref{tab:MulRegSONA2022}. The current model is shown to be a strong predictor of tag attitude variance ($R^2$ of 0.38 and adjusted $R^2$ of 0.30). Political orientation is the most influential variable explaining 39.01\% of variance for tag attitudes as shown from the Shapley results. Further, political liberalism was positively associated with tag attitudes when controlling for all other tested variables ($\beta$ = -0.28, p < 0.001). BFI Openness was the second most influential variable contributing up to 19.18\% of tag attitude variance, and was shown to be positively associated when controlled for all other factors ($\beta$ = 0.21, p < 0.05). Trust in Medical Scientists was the third most influential variable contributing to 13.54\% of tag attitude variance. This model also shows that Trust in Medical Scientists is positively correlated with tag attitudes when controlled for all other traits and demographic factors ($\beta$ = 0.15, p < 0.05). No other variables in the model showed a significant effect with aggregate tag attitude. 

\begin{table*}
  \small
  \caption{\centering Multiple regression - Tag attitude aggregate as dependent variable (SONA)}
  \label{tab:MulRegSONA2022}
  \begin{tabular}{p{2.3cm}|p{1.2cm}|p{1.4cm}|p{1.0cm}|p{1.0cm}|p{1.0cm}}
    \toprule
    & \textbf{$\beta$ coef} & \textbf{Shapley $R^2$} & \textbf{std err} & \textbf{t-value} & \textbf{p-value}\\
    \midrule
    \textbf{Politics} & \textbf{-0.28} & \textbf{39.01\%} & \textbf{0.07} & \textbf{-4.05} & \textbf{0.00}\\
    CRT & -0.01 & 0.30\% & 0.04 & -0.34 & 0.73 \\
    NFCC & 0.09 & 1.95\% & 0.09 & 1.08 & 0.28 \\
    \textbf{Openness} & \textbf{0.21} & \textbf{19.18\%} & \textbf{0.10} & \textbf{2.00} & \textbf{0.05} \\
    Conscientiousness & 0.15 & 5.61\% & 0.09 & 1.55 & 0.12 \\
    Agreeableness & 0.08 & 7.95\% & 0.11 & 0.77 & 0.44 \\
    Extraversion & -0.03 & 0.91\% & 0.10 & -0.32 & 0.75 \\
    Neuroticism & 0.09 & 2.72\% & 0.08 & 1.21 & 0.23 \\
    Trust Elected & 0.07 & 0.86\% & 0.08 & 0.94 & 0.35 \\
    Trust News Media & -0.01 & 0.34\% & 0.07 & -0.11 & 0.91 \\
    \textbf{Trust Med Sci} & \textbf{0.15} & \textbf{13.54\%} & \textbf{0.07} & \textbf{2.24} & \textbf{0.03} \\
    Trust Religious & -0.07 & 7.21\% & 0.07 & -1.00 & 0.32 \\
    Age & 0.00 & 0.42\% & 0.01 & 0.02 & 0.98\\
    \midrule
    \textbf{$R^2$} & 0.38 & & & \\
    \textbf{Adj $R^2$} & 0.30 & & & \\
    \bottomrule
  \end{tabular}
\end{table*}

Regression results for the MTurk sample are shown in Table \ref{tab:mulRegMTURK2021}. Within this model, NFCC is the most influential variable explaining up to 59.77\% of variance for the tag attitude scale. Further, NFCC is positively correlated with favorable tag attitudes and is statistically significant when controlled for all other factors ($\beta$ = 0.29, p < 0.001). The second most influential variable, Trust in Medical Scientists, is also positively correlated with tag attitudes ($\beta$ = 0.11, p<.01) and explains 9.94\% of model variance. Conscientiousness, conservatism, and trust in religious figures are all negatively correlated with tag attitudes ($\beta$ = -0.18, p < 0.01;$\beta$ = -0.05, p < 0.05; = -0.09, p < 0.05). The total $R^2$ of the model is 0.37 and the adjusted $R^2$ is 0.32. 

\begin{table*}
  \small
  \caption{\centering Multiple regression - Tag attitude aggregate as dependent variable (MTurk)}
  \label{tab:mulRegMTURK2021}
  \begin{tabular}{p{2.3cm}|p{1.2cm}|p{1.4cm}|p{1.0cm}|p{1.0cm}|p{1.0cm}}
    \toprule
    & \textbf{$\beta$ coef} & \textbf{Shapley $R^2$} & \textbf{std err} & \textbf{t-value} & \textbf{p-value}\\
    \midrule
    \textbf{Politics} & \textbf{-0.05} & \textbf{7.78\%} & \textbf{0.02} & \textbf{-2.35} & \textbf{0.02}\\
    CRT & 0.06 & 2.01\% & 0.04 & 1.54 & 0.13\\
    \textbf{NFCC} & \textbf{0.29} & \textbf{59.77\%} & \textbf{0.04} & \textbf{6.59} & \textbf{0.00}\\
    \textbf{Conscientiousness} & \textbf{-0.18} & \textbf{9.18\%} & \textbf{0.07} & \textbf{-2.49} & \textbf{0.01}\\
    Trust Elected & 0.02 & 1.46\% & 0.05 & 0.42 & 0.68\\
    Trust News Media & 0.08 & 4.99\% & 0.05 & 1.74 & 0.08\\
    \textbf{Trust Med Sci} & \textbf{0.11} & \textbf{9.94\%} & \textbf{0.04} & \textbf{2.52} & \textbf{0.01}\\
    \textbf{Trust Religious} & \textbf{-0.09} & \textbf{3.80\%} & \textbf{0.04} & \textbf{-1.99} & \textbf{0.05}\\
    Age & 0.00 & 1.07\% & 0.00 & -0.62 & 0.53\\
    \midrule
    \textbf{$R^2$} & 0.37 & & & \\
    \textbf{Adj $R^2$} & 0.32 & & & \\
    \bottomrule
  \end{tabular}
\end{table*}

\subsection{SONA and MTurk Comparison}
Though SONA and MTurk showed correlations with respect to Conscientiousness in the bivariate correlation, they did not share any other correlations with respect to the aggregate tag attitude - though they did with certain individual attitude measures. While NFCC showed no significant overall effects for SONA, among MTurkers NFCC was positively correlated with the aggregate tag attitude scale (r = 0.447, p < 0.01) as well as all individual items. Significant effects found in the SONA sample but not in the MTurk sample include Politics, Trust in Medical Scientists, and Trust in Religious Leaders. BFI Openness and Agreeableness were not tested in the MTurk sample. 

Comparing multiple regression results, both found Politics and Trust in Medical Scientists to be significant predictors of attitude when controlling for the other tested variables. The SONA sample also found Openness to be significant, which was not data available in the MTurk sample. Further, the MTurk sample found NFCC, Conscientiousness, and Trust in Religious Leaders to be statistically significant while the SONA sample did not. It is possible that the SONA regression model may show fewer significant effects than MTurk due to having a greater number of examined covariates. 

\subsection{Summary of Results}
In summary, the bivariate correlations for the SONA sample indicated that Conscientiousness, Openness, Agreeableness, Politics, Trust in Medical Scientists, and Trust in Religious Leaders were predictors of tag attitudes. When controlling for covariates in the multiple regression, the effects from Politics, Openness, and Trust in Medical Scientists remained significant (p < 0.05) in the SONA sample. Bivariate correlation results of the MTurk sample was significant for both NFCC (p < 0.01) and Conscientiousness (p < 0.05). These effects remain significant when controlling for all tested variables in the MTurk sample, with NFCC being the strongest predictor by a large margin. Politics, Trust in Medical scientists, and Trust in Religious Figures also showed significant associations in the multiple regression model, despite not having significant bivariate correlations with the tag attitude scale. This indicates that the effects from Politics, Trust in Medical Scientists and Religious Figures are not as robust compared to the other tested factors in the MTurk sample. While some traits showed no significant effects with the aggregate scale, some showed significant correlations with individual attitude items that could still have implications for future warning tag design. Differences between the two distinct samples tested in this study further emphasize the importance of individual differences in understanding the impact of misinformation mitigation strategies.

\section{Discussion}
The study presented here supports the premise that cognitive factors, beliefs, and personality traits impacts how a person views and interacts with misinformation warning tags. Specifically, our results indicate that several individual factors - including Politics, NFCC, Openness, Conscientiousness, Agreeableness, Trust in Medical Scientists, and Trust in Religious Leaders - predict a person's attitudes towards misinformation warning tags. When controlling for effects from covariates and multicollinearity, Politics, Openness, and Trust in Medical Scientists were the strongest predictors for the SONA sample while NFCC explained the majority of tag attitude variance among MTurk workers. These results may explain discrepancies in past literature on the effectiveness of tags as a mitigation strategy, as certain tags may work for some people but not others. Our results highlight the premise that designing tailored misinformation interventions as opposed to taking one-size-fits-all strategies may benefit designers seeking to curb misinformation for diverse groups.

Predictably, we find differences in the most important characteristics between samples composed of different populations of people. Differences seen between the samples strengthens the case that different populations may have different needs for warning tags, and thus may require different types of personalization. Improved predictions of a user's attitude towards warning labels can be leveraged by future designers as a means to personalize interventions that work best for specific user groups. In the remainder of this paper, we will discuss the implications of our results and provide recommendations for designing more effective misinformation warning tags.

\subsection{Individual Factors Predict Warning Tag Attitudes}
\textbf{Information assessment traits may be strong predictors of tag attitudes.} We found that MTurk workers who have a desire for certainty and predictability (i.e. high NFCC) have positive attitudes towards warning tags. Intuitively, this makes sense: warning tags have the potential to bring clarity as to whether or not a post's information can be relied upon. Interestingly, we did not find significant effects between NFCC and tag attitudes for the SONA sample. This distinction between the SONA and MTurk sample warrants further investigation, though we suspect it may be due to the overall higher tolerance for uncertainty observed by the university students as compared to our more general MTurk population (see Appendix Table \ref{tab:meanCompPlat} for mean comparisons).

Though Cognitive Reflective Test (CRT) score did not show significant aggregate results, individual items “I am more likely to research the post’s topic in order to evaluate the content myself" (positive correlation) in the SONA sample and "I usually ignore the tag" (negative correlation) in the MTurk sample validates prior work that respondents higher in Cognitive Reflective Test (CRT) score may take behaviors that allow them be more capable of distinguishing real from fake news \cite{Pehlivanoglu2020TheRO}, potentially due to their ability and propensity to research the topics themselves once signaled that a post may contain misinformation.

\vspace{1em}
\noindent
\textbf{Political orientation is one indicator of warning tag attitudes, but does not show the whole picture.} We echo prior study results showing Politics to play a large role in attitudes towards warning tags \cite{Saltz2021MisinformationIA}. Indeed, Politics was the strongest predictor of tag attitudes for SONA. For MTurk, however, Politics was a weaker predictor of tag attitudes compared to other significant effects. In both samples, liberal leaning respondents were more likely to have a positive attitude towards warning tags while conservative leaning respondents were were more likely to have negative attitudes. 

Importantly, while Politics may be an effective indicator of tag attitudes in some cases, our findings indicate that other factors such as Openness and NFCC are also strong predictors of a respondent's attitude towards warning tags. The inclusion of these other factors can provide a more comprehensive understanding of how different groups may think about and behave towards warning tags. In a more general sense, these results imply that the research community should look beyond more superficial factors like political orientation and examine other personal characteristics - including a person's cognitive and psychological profile and trust judgments - to to deepen our understanding of how they may interact with misinformation mitigation strategies like tags. 

\vspace{1em}
\noindent
\textbf{Personality matters: BFI Openness was positively associated with tag attitudes for SONA, while Conscientiousness was negatively associated for Mturk} when controlling for tested covariates. When considering results from the bivariate correlations, Agreeableness was also  positively correlated with tag attitudes. Our findings show personality is an important predictor of how a person thinks and behaves towards misinformation warning tags.

We found that undergraduates who are more receptive to new ideas - i.e. are more Open - have more favorable attitudes towards warning tags. This is consistent with previous work showing that Openness is associated with better news discernment and skepticism towards myths \cite{calvillo2020political, swami2016believes}. We believe openness in this case may be reflective of how willing a person is to accept an outside judgment of the credibility of information. It is important to keep in mind that Openness was not tested in the MTurk sample, and thus comparisons cannot be drawn between samples.

Our results indicate that respondents who are more Agreeable - i.e. they have a greater care for social harmony and cohesion - may have more positive attitudes towards warning label interventions. This intuitively makes sense as well: misinformation warning tags are, in a general sense, pro-social, and may be seen as a way to promote truth, show care for one's community, and reduce the negative impacts of spreading false information. This aligns with a previous work claiming users high in agreeableness are more likely to validate news prior to sharing said news \cite{sampat2022fake}. Importantly, Agreeableness was positively correlated with aggregate tag attitude in the bivariate SONA analysis, but not the multiple regression model. This implies that agreeableness may be an important factor to consider, but perhaps less important that other characteristics.

MTurk respondents who are more self-disciplined, organized, and responsible - i.e. Conscientious - are more likely to have negative attitudes towards warning tags. This could stem from a desire for self-sufficiency, a confidence in their own discernment abilities, or a preference to avoid outside judgments if they believe they can handle the task themselves. If the less favorable views of tags among Conscientious individuals is driven by stronger confidence in information assessment abilities, then the present results may still be consistent with previous work showing associations between Conscientiousness with better news discernment and lower tendency to share misinformation \cite{calvillo2020political, lawson2022pandemics}. We did find mixed results, however, when comparing samples: Conscientiousness was negatively correlated with aggregate tag attitudes in the MTurk sample, but showed a positive correlation for SONA respondents. It is also worth noting that the effect from Conscientiousness from MTurkers remained significant when controlled for the influence from the other tested variables while it no longer showed an effect in the SONA regression model. This may indicate that for MTurk workers, Conscientiousness is a more influential trait for misinformation-related outcome measures. This is further suggested in related work, which showed that Conscientiousness was the strongest predictor for misinformation detection accuracy among MTurkers \cite{Kaufman2022WhosIT}. The difference in the direction of correlation effects may also be attributed to contextual differences when recruiting respondents from MTurk compared to SONA. For MTurk workers, Conscientiousness may be a more influential factor because being orderly and cautious would be more useful attributes when completing other tasks on the platform, which typically involve an attention to detail to receive approval for the work (e.g., correctly labeling images to be used as training data). In contrast, undergraduates enrolled in a university course may place less importance on being Conscientious, as they are not receiving monetary compensation for completing tasks requiring attention to detail. Undergraduates may also have more favorable views of social media companies in general, which could result in more favorable views of tagging interventions. Results in Table \ref{tab:bivCorSONA2022} support this possibility, where Conscientious was positively correlated with the statement "I appreciate the social media platform's attempt to warn me about potential misinformation" among SONA respondents. 

\vspace{1em}
\noindent
\textbf{The sources of information a person trusts may be an important predictor of attitude.} Trust in Medical Scientists showed strong, positive association with warning tag attitudes in the SONA sample, meaning respondents who had greater Trust in Medical Scientists were more likely to have positive attitudes about warning tags. As prior work shows that those who use factual news sources have higher trust in medical scientists \cite{Larsen2023TheIO}, the positive associations with tag attitudes may reflect a higher respect for institutional guidance and public health recommendations. More generally, it may also reflect a person's preference for expert judgment and evidence-based decision-making. 

Conversely, Trust in Religious Leaders was found to be negatively correlated with warning tag attitudes in the SONA sample. One explanation for this finding may be due to differences in the criteria by which credibility of information is assessed: people higher in Trust in Religious Leaders may prioritize spiritual authorities over scientific authorities. A second hypothesis suggests the finding may stem from a distrust in secular institutions like social media platforms. This latter point is supported by the individual measure finding showing people high in Trust in Religious Leaders had a higher distrust in social media platform interventions. Though prior work has found that religious beliefs to not indicate conspiracy theory belief \cite{JasinskajaLahti2019UnpackingTR}, differences in credibility assessment and trust in institutions has important implications on misinformation intervention design. In the MTurk sample, responses the items "I usually ignore the tag" and “I am more likely to research the post’s topic in order to evaluate the content myself" might indicate that respondents high in Trust in Religious Leaders likely don't trust social media platforms, are using a different set of evaluation criteria to determine truth in content, or are skeptical of interventions in general. 

For Trust in News Media in the MTurk sample, the meaningfulness of the response "I appreciate the social media platform’s attempt to warn me about potential misinformation" may indicate respondents that trust news media also trust social media platforms. In both the SONA and MTurk sample, the aggregate was insignificant.

\subsection{Design Implications}
In this study, we found that certain traits can predict people's attitudes and self-described behaviors towards misinformation warning tag interventions. These results have implications for the design of misinformation mitigation solutions like warning tags. Most importantly, our results highlight the need for personalized and culturally-relevant warning tag interventions, as one-size-fits-all approaches to prevent misinformation uptake and spread may fail if they don’t harmonize with the specific needs and motivations of a particular individual. These include structuring messages for audiences with different psychological, cognitive, and information-assessment traits as well as using an understanding of credibility signaling to help people who trust different authority sources heed similar warnings. Taken together, our findings provide actionable direction for future warning tag design and misinformation mitigation solution design more generally.

\vspace{1em}
\noindent
\textbf{Designing For Cognitive and Informational Assessment Traits.}
Our results imply that cognitive and information assessment traits like Need for Cognitive Closure (NFCC) and Cognitive Reflective Test (CRT) scores may meaningfully impact a person's attitude and behavior towards misinformation warning tags. The general implication is that these types traits should be accounted for in design of warning tags and other misinformation mitigation solutions. The results of our study can provide some direction for designers, however, further research is needed on how exactly to design for people with high and low NFCC and CRT. As we may not always be able to detect a person's cognitive and informational assessment traits, it may be best to design for the anti-tag attitude cases.

People high in NFCC prefer clear, unambiguous information, and thus we saw in our study that these individuals (particularly in our MTurk sample) had more favorable attitudes towards warning tags, and thus warning tags may be appropriate and effective for this population. A more challenging design problem is designing for people \textit{low} in NFCC. For these people, messaging regarding the importance of certainty and clarity on the truthfulness of information and the implications of spreading false information may be helpful. In this way, even people who do not have the tendency to care about ambiguity may engage in more information discernment behaviors. 

Our results suggest that people high in CRT are less likely to ignore the warning (MTurk) and are more likely to research the topic themselves (SONA). Warning tag interventions may capitalize on the user's interest in self determination by presenting relevant research articles or news from reputable sources. For misinformation in the health domain specifically, harder moderation approaches with detailed explanation on why the determination was made may be appropriate for these users. 

Users lower in CRT may warrant behavioral intervention or harder ("hands-off") moderation approaches. Since these users tend to ignore warning tags and are less likely to conduct their own research, prompts requesting behavior (for example, \textit{"Are you sure you want to share this potentially false post?"}) may be helpful. Severe cases of misinformation may need to be completely removed for these users.

\vspace{1em}
\noindent
\textbf{Designing For Personality.}
Our results indicate that tailoring misinformation warning tags and other mitigation solutions based on a person's personality may be an effective approach. Prior work has shown connections between personality and misinformation vulnerability, and our study supports this direction of research \cite{Kaufman2022WhosIT, Lai2020WhoFF} and spreading behavior \cite{Indu2021ASR, Alvi2020INFORMATIONP, lawson2022pandemics}. Recent work in social media research has shown that personality can be accurately detected from a person's social media behavior \cite{farnadi2016computational} - we suggest that similar approaches may be used as a means to inform the tailoring of personality-based warning tags.

People high in Openness may benefit from warning tags as a means to show them external perspectives. To design for people low in Openness, messaging emphasizing the importance of exploring new, diverse, or more objective perspectives to their own may be necessary for warning tag acceptance. Considering people low in Openness are less interested in conducting their own research, providing justification for misinformation as part of the label may provide a prudent safe guard for these more susceptible users. Other priming interventions, such as "accuracy prompts" which asks users to rate a neutral headline to encourage users to scrutinize future articles \cite{pennycook2021psychology}, may also be pertinent. 

Similar to Openness, people high in Agreeableness may benefit from warning tags inherently. For people high in Agreeableness, emphasis on the pro-social nature of warnings may increase their effectiveness. For people low in Agreeableness, prompts to increase empathy, orient the person to understand the impact of misinformation on vulnerable communities, and emphasize compassion may be helpful to increase warning tag efficacy. 

People high in Conscientiousness may benefit from warning tags emphasizing the importance of sharing verified information, whereas those low in Conscientiousness may be less organized, so simple messaging and clear visuals that require less detail orientation may be helpful. Those high in Conscientiousness and more trusting of social media platforms (such as from our undergraduate sample) will likely benefit from warning tag interventions as they are. Users similar to those in the MTurk population that are also high in Conscientiousness trust warning tags less and may require interventions that increase trust for these particular users. Providing explanations about the underlying decision making process (algorithmic, community, 3rd party, etc.), could increase user trust with platforms as well. We posit 3rd party moderators may be particularly effective at increasing platform trust with these users.

\vspace{1em}
\noindent
\textbf{Designing For Trust And Credibility Based On Identity And Belief.}
Our study results suggest that the informational sources one trusts as well as their belief systems impact a person's attitudes towards misinformation warning tags.

In the case of political orientation, self-identified democrats were generally more positive about warning tags than republicans. Understanding \textit{why} there may be differences based on political ideology may reflect differences in media consumption habits, including trust in mainstream media and the perceived bias of misinformation-identifying algorithms, as well as the bias of the misinformation itself (i.e. if the misinformation supports republican or democratic-led claims). Designing warning tags based on partisan divides is a challenge when one wants to remain objective in the discernment of trust based on demonstrable fact, however, some general principles may apply. There is evidence to suggest that republicans and democrats may differ in their value judgments \cite{haidt2012righteous}, and thus tailored tags that appeal to each group's value systems may be a good place to start. For example, democrats may respond better to appeals to institutional authority, collective well-being, or scientific argument, whereas republicans may respond better to appeals to individual freedom and personal responsibility. In both cases, endorsement by authority figures from their respective parties may help increase the effectiveness of warning tags.

In a similar vein, we find that the specific institutions a person trusts impacts their warning tag attitudes. Our study found people high in Trust in Medical Scientists had more favorable tag attitudes, while people with high Trust in Religious Leaders had less favorable tag attitudes. These imply the importance of appealing credibility to the person's authority figures of choice as well as messaging that aligns with a person's reasoning priorities. For example, warning tags for people high in Trust in Medical Scientists may benefit from logical arguments based in scientific reasoning, while people low in Trust in Medical Scientists but high in trust in a different institution - such as Religious Leaders - may benefit from appeals to specific spiritual values, culturally-specific messaging, and moral responsibility. Previous work showing that medical authority is often used to propagate misinformation (e.g., stating a false health treatment is endorsed by a medical professional) \cite{haupt2021identifying, haupt2023detecting} should also be taken into consideration when designing tag interventions, as this increases the difficulty for algorithms to detect misinformation. In these scenarios, tag labels that hedge or use less certain language (e.g., this may contain misinformation) may be more effective by mitigating the effects from potentially labeling accurate scientific information as misinformation. For people high in Trust in Religious Leaders specifically, however, it is possible that warning tags may not be the best intervention. These users may be more likely to ignore tags and may even have \textit{more} positive views of authors of potential misinformation. More generally, interventions may also attempt to improve trust by catering to a person's preference of moderation group (community, third party, algorithmic) to best determine what should be removed, as long as it remains objective. 

\vspace{1em}
\noindent
\textbf{Broader Implications.}
Up until this point, we have primarily focused on individual differences with regards to warning tag interventions for online misinformation specifically. We expect that many of the implications of this study - particularly the need for personalization and culturally-relevant design - to generalize to other types of misinformation interventions or even to other domains where algorithmically-determined decisions need to be trusted or understood by a particular user, such as explainable AI. In particular, we expect that many of the design implications detailed above, including how to design for cognitive and informational traits, personality, and beliefs about trust and credibility, may generalize to other contexts. The concept of personalization not new - personalization and tailoring has been commonplace in public health and advertising for decades \cite{kreuter2013tailoring}, and has become a major topic in human-computer (particularly human-AI) interaction \cite{holzinger2019causability, wang2019designing, schneider2019personalized, kaufman2022cognitive, pazzani2022expert}. Our results support this direction of work and may be used by practitioners in a variety of HCI domains as a means to better understand and design for diverse populations.

\section{Limitations and Future Work}
This study was not without limitations. This study was conducted in the United States, as such the results may not generalize to other populations. The SONA results may be more applicable to younger/liberal leaning populations, while MTurk results may can be applied to more general populations. MTurk workers were motivated via financial incentives while SONA participants were motivated by class credits, resulting in potentially different values when taking the survey, and though unlikely, it is possible compensation may have biased results. There is always the risk of dishonest answers from respondents who partook in this study, though the study was voluntary and anonymous to limit this risk. Since data is self-reported, we do not know how actual behavior in real world settings may differ from the reported attitudes in this study.

These limitations motivate potential future research and design studies. A user study testing the effectiveness of tag interventions "in the wild" to better asses real world behavior towards different types of tailored warning tags would be a natural next step. As our results imply differences based on individual traits, future work may sample from populations in other countries to assess cultural differences in warning tag attitudes; warning tags may then be personalized accordingly to these additional user populations. Beyond cultural differences, future work may investigate additional populations or characteristics to better understand how attitudes may differ.

\section{Conclusion}
In this study, we investigated how cognitive factors, beliefs, and personality impact attitudes towards misinformation warning tags. This is an important area of study, as online misinformation is rampant and warning tags may provide an effective mitigation solution. Using two distinct sample groups, we find that Openness, Agreeableness, NFCC, CRT, and Trust in Medical Scientists are positively correlated with attitudes towards warning tag and Trust in Religious Leaders, Conscientiousness, and political conservatism are negatively correlated with attitudes towards warning tags. These characteristics can be used to predict user attitudes towards warning labels and be leveraged as a means to personalize misinformation warning tags as a means to potentially increase their efficacy for diverse populations. Results are synthesized into design considerations which provide specific direction for how to design online misinformation interventions based on a person's cognitive and informational traits, personality, and beliefs about trust and credibility. The implications of this study may generalize to other domains where algorithmically-determined decisions are deployed at-scale, including other human-computer interaction domains. 

\begin{acks}
The authors would like to acknowledge Steven Dow, Chloe Lee, and Joseline Chang for supporting the data collection process. A special thanks to the Discovery Way Foundation for their generous usage of lab space.
\end{acks}

\bibliographystyle{ACM-Reference-Format}
\bibliography{bibliography}
\clearpage

\appendix
\section{Additional Tables}
\subsection{Mean Comparison of MTurk and SONA}
Based on Table \ref{tab:meanCompPlat}, the MTurk sample is older and more conservative compared to SONA. MTurkers also show a statistically significant difference with higher Trust in News Outlets, Trust in Religious Leaders, Trust in Elected Officials, and NFCC compared to SONA respondents. MTurkers also have slightly more favorable attitudes towards warning tags on average than SONA.

\begin{table}[h!]
\centering
\caption{Mean Comparisons in Factor Measures by Platform}
\label{tab:meanCompPlat}
\begin{tabular}{lcccccc}
\toprule
& \textbf{MTurk} & \textbf{SONA} & \textbf{Diff} & \textbf{\textit{Scale Range}} \\
\midrule
\textbf{Age} & \textbf{37.25} & \textbf{21.63} & \textbf{15.62} & Open \\
\textbf{Political Conservativeness} & \textbf{3.39} & \textbf{2.54} & \textbf{0.85} & (1 to 6) \\
\textbf{Trust News} & \textbf{3.09} & \textbf{2.04} & \textbf{1.05} & (1 to 4) \\
Trust Medical & 3.07 & 3.13 & -0.06 & (1 to 4) \\
\textbf{Trust Religious} & \textbf{2.69} & \textbf{1.65} & \textbf{1.04} & (1 to 4) \\
\textbf{Trust Elected} & \textbf{2.85} & \textbf{1.83} & \textbf{1.02} & (1 to 4) \\
\textbf{NFCC} & \textbf{4.29} & \textbf{3.97} & \textbf{0.32} & (1 to 6) \\
CRT & 0.98 & 1.15 & -0.17 & (0 to 3) \\
Conscientious & 3.34 & 3.44 & -0.10 & (1 to 5) \\
Extraversion & NA & 3.08 & NA & (1 to 5) \\
Openness & NA & 3.74 & NA & (1 to 5) \\
Neuroticism & NA & 3.12 & NA & (1 to 5) \\
Agreeableness & NA & 3.46 & NA & (1 to 5) \\
\textbf{Aggregate Tag Score} & \textbf{3.60} & \textbf{3.46} & \textbf{-0.1} & (1 to 5) \\
\bottomrule
\end{tabular}
\begin{tabular}{@{}l@{}}
\textit{Note: Emboldened characteristics are significantly different }\\
\textit{at p < 0.05 in the two-sided test of equality for column means.}
\end{tabular}
\end{table}

\end{document}